\documentclass[12pt]{article}
\usepackage{amsmath,amssymb}

\global\arraycolsep=1pt
\numberwithin{equation}{section}
\newcommand{\bel}[1]{\begin{equation}\label{#1}}                     
\newcommand{\bal}[1]{\begin{eqnarray}\label{#1}}                     
\newcommand{\be}{\begin{equation}}
\newcommand{\ee}{\end{equation}}

\newcommand{\ex}{\mathrm{e}}
\newcommand{\de}{d}

\newcommand{\scr}{\scriptstyle}
\newcommand{\qq}{\qquad}

\renewcommand{\thefootnote}{\fnsymbol{footnote}}

\begin{document}


\begin{flushright}
July, 2007 \\
OCU-PHYS 271 \\
\end{flushright}
\vspace{20mm}

\begin{center}
{\bf\Large
Closed conformal Killing-Yano tensor\\
and geodesic integrability
}
\end{center}

\begin{center}

\vspace{15mm}

Tsuyoshi Houri$^a$\footnote{
\texttt{houri@sci.osaka-cu.ac.jp}
}, 
Takeshi Oota$^b$\footnote{
\texttt{toota@sci.osaka-cu.ac.jp}
} and
Yukinori Yasui$^a$\footnote{
\texttt{yasui@sci.osaka-cu.ac.jp}
}

\vspace{10mm}

${}^a$
\textit{
Department of Mathematics and Physics, Graduate School of Science,\\
Osaka City University\\
3-3-138 Sugimoto, Sumiyoshi,
Osaka 558-8585, JAPAN
}
\vspace{5mm}

\textit{
${{}^b}$
Osaka City University
Advanced Mathematical Institute (OCAMI)\\
3-3-138 Sugimoto, Sumiyoshi,
Osaka 558-8585, JAPAN
}

\vspace{5mm}

\end{center}
\vspace{8mm}

\begin{abstract}
Assuming the existence of a single rank-$2$ closed conformal Killing-Yano
tensor with a certain symmetry we show that there exist mutually commuting
rank-$2$ Killing tensors and Killing vectors.
We also discuss the condition of separation of variables
for the geodesic Hamilton-Jacobi equations.
\end{abstract}

\vspace{25mm}

\newpage

\renewcommand{\thefootnote}{\arabic{footnote}}
\setcounter{footnote}{0}


\section{Introduction}

Recently, it has been shown that geodesic motion in the
Kerr-NUT de Sitter spacetime is integrable 
for all dimensions \cite{FK,KF,PKVK,FKK,KKPF,KKPV}. 
Indeed, the constants of motion that are in involution
can be explicitly constructed from a rank-2 closed conformal
Killing-Yano (CKY) tensor. In this paper we consider the problem of integrability
of the geodesic equation in a more general situation. We
assume the existence of a single rank-2 closed CKY tensor with a certain symmetry 
for $D$-dimensional spacetime $M$ with a metric $g$.
It turns out that such a spacetime admits mutually commuting $k$ rank-2 Killing tensors
and $k$ Killing vectors. Here we put $D=2 k$ for even $D$ , and $D=2 k-1$ for odd $D$.
Although the existence of the commuting Killing tensors was 
shown in \cite{KKPF,KKPV},  we reproduce it more directly. 
We also discuss the condition of separation of variables for
the geodesic Hamilton-Jacobi equations using the result given 
by Benenti-Francaviglia \cite{BF} and Kalnins-Miller \cite{KM} (see also \cite{ben}).

\section{Assumptions and main results}

A two-form 
\be
h = \frac{1}{2} h_{ab}\, \de x^a \wedge \de x^b, \qq
h_{ab} = - h_{ba}
\ee
is called a conformal Killing-Yano~(CKY) tensor
if it satisfies
\begin{equation} \label{CKY}
\nabla_a h_{bc}+\nabla_b h_{ac}= 2 \xi_c g_{ab}- \xi_a g_{bc}-\xi_b g_{ac}.
\end{equation} 
The vector field $\xi_a$ is called
the associated vector of $h_{ab}$, which is given by
\be
\xi_a = \frac{1}{D-1} \nabla^b h_{ba}.
\ee
In the following we assume
\begin{equation}
(a1)~dh=0,~~~
(a2)~\mathcal{L}_{\xi} g=0,~~~
(a3)~\mathcal{L}_{\xi} h=0.
\end{equation}
The assumption $(a1)$ means that $(D-2)$-form $f= \ast h$ is a Killing-Yano~(KY)
tensor,
\begin{equation}
\nabla_{(a_1} f_{a_2)a_3 \cdots a_{D-1}}=0.
\end{equation}
Note that the equation \eqref{CKY} together with $(a1)$ is equivalent to 
\begin{equation} \label{CCKY}
\nabla_a h_{bc}= \xi_c g_{ab}-\xi_b g_{ac}.
\end{equation} 
It was shown in \cite{tac} that the associated vector $\xi$ satisfies
\begin{equation}
\nabla_a \xi_b+ \nabla_b \xi_a=\frac{1}{D-2}(R_{a}{}^{c}\, h_{bc}+R_{b}{}^{c}\, h_{ac}),
\end{equation} 
where $R_{ab}$ is a Ricci tensor. If $M$ is Einstein,~i.e. $R_{ab}= \Lambda g_{ab}$,
then
\begin{equation} 
\nabla_a \xi_b+ \nabla_b \xi_a=0.
\end{equation}
Thus, any Einstein space satisfies the assumption $(a2)$ \cite{tac}.
According to \cite{KKPF}, we define 2$j$-forms $h^{(j)}$~($j=0,\cdots,k-1$):
\be
h^{(j)} = \underbrace{h \wedge h \wedge \dotsm \wedge h}_{j}
= \frac{1}{(2j)!} h^{(j)}_{a_1 \dotsm a_{2j}}
\de x^{a_1} \wedge \dotsm \wedge \de x^{a_{2j}},
\ee
where the components are written as
\be
h^{(j)}_{a_1 \dotsm a_{2j}}
= \frac{(2j)!}{2^j} h_{[a_1 a_2} h_{a_3 a_4} \dotsm h_{a_{2j-1} a_{2j}]}.
\ee
Since the wedge product of two CKY tensors is again a CKY tensor,
$h^{(j)}$ are closed CKY tensors, and so
 $f^{(j)}= \ast h^{(j)}$ KY tensors. 
Explicitly, we have
\be
f^{(j)} = * h^{(j)} = \frac{1}{(D-2j)!}
f^{(j)}_{a_1 \dotsm a_{D-2j}}
\de x^{a_1} \wedge \dotsm \wedge \de x^{a_{D-2j}},
\ee
where
\be
f^{(j)}_{a_1 \dotsm a_{D-2j}}
= \frac{1}{(2j)!} \varepsilon^{b_1 \dotsm b_{2j}}{}_{a_1 \dotsm a_{D-2j}}
h^{(j)}_{b_1 \dotsm b_{2j}}.
\ee
Given these KY tensors, we can construct the rank-2 Killing tensors $K^{(j)}$ obeying
the equation
$
\nabla_{(a}K^{(j)}_{bc)}=0$~:
\bel{DKj}
K^{(j)}_{ab}= \frac{1}{(D-2j-1)!(j!)^2}
f^{(j)}_{a c_1 \dotsm c_{D-2j-1}} f_b^{(j)c_1 \dotsm c_{D-2j-1}}.
\ee 
From $(a2)$ we have 
$\mathcal{L}_{\xi} \ast\! h^{(j)}=\ast\, \mathcal{L}_{\xi} \, h^{(j)}$ 
and hence
the assumption $(a3)$ yields
\begin{equation} \label{LieD}
\mathcal{L}_{\xi} h^{(j)}=0,~~\mathcal{L}_{\xi} f^{(j)}=0,~~ \mathcal{L}_{\xi} K^{(j)}=0.
\end{equation}
We also immediately obtain from \eqref{CCKY}
\begin{equation} \label{CovD}
\nabla_{\xi} h^{(j)}=0,~~\nabla_{\xi} f^{(j)}=0,~~ \nabla_{\xi} K^{(j)}=0.
\end{equation}

Let us define the vector fields $\eta^{(j)}$ by \cite{DF,DR}
\begin{equation}
\eta^{(j)}_a = K^{(j)}{}_{a}{}^b\xi_b. 
\end{equation}
Then we have
\begin{equation}
\nabla_{(a}\eta^{(j)}_{b)} =
\frac{1}{2}\mathcal{L}_{\xi}K^{(j)}_{ab}-\nabla_{\xi}K^{(j)}_{ab},
\end{equation}
which vanishes by \eqref{LieD} and \eqref{CovD},~i.e. $\eta^{(j)}$ are Killing vectors. 

Theorem $1$ was proved
in \cite{KKPF,KKPV}.\\
{\bf{Theorem 1}} ~~Under $(a1)$ Killing tensors $K^{(i)}$
 are mutually commuting,
\[ [K^{(i)},~ K^{(j)}]_S =0. \]
The bracket $[~~ , ~~]_S$ represents a symmetric Schouten product. The equation can be
written as
\begin{equation} \label{Thm1}
K^{(i)}_{d(a} \nabla^{d} K^{(j)}_{bc)}-K^{(j)}_{d(a} \nabla^{d} K^{(i)}_{bc)}=0.
\end{equation}
Adding the assumptions $(a2)$ and $(a3)$ we prove\\
{\bf{Theorem 2}}~~
\be
\mathcal{L}_{\eta^{(i)}} h = 0.
\ee
{\bf{Corollary}}~~
Killing vectors $\eta^{(i)}$ and Killing tensors $K^{(j)}$
are mutually commuting,
\[ [\eta^{(i)},~ K^{(j)}]_S =0,~~[\eta^{(i)},~ \eta^{(j)}] =0. \]


\section{Proof of theorems 1,2 }

Let $H$, $Q:=-H^2$, $K^{(j)}$ be matrices with elements
\be
H^a{}_b = h^a{}_b,
\qq
Q^{a}{}_b = - h^a{}_c h^c{}_b,
\qq
(K^{(j)})^a{}_b = K^{(j)a}{}_b.
\ee
The generating function of $K^{(j)}$ can be read off from \cite{KKPF}:
\bel{GFK}
K_{ab}(\beta) = \sum_{j=0}^{k-1} K^{(j)}_{ab} \beta^j
= \det\!{}^{1/2} ( I + \beta Q)\, \Bigl[ ( I + \beta Q)^{-1}
\Bigr]_{ab}.
\ee
Here $k=[(D+1)/2]$.
Note that
\be
\begin{split}
& 2 \det\!{}^{1/2} ( I + \beta Q)\, \Bigl[ ( I + \beta Q)^{-1}
\Bigr]^{a}{}_{b} \cr
&= \det( I + \sqrt{\beta} H ) \, 
\Bigl[ ( I + \sqrt{\beta} H )^{-1} \Bigr]^a{}_b
+ \det( I - \sqrt{\beta} H ) \, 
\Bigl[ ( I - \sqrt{\beta} H )^{-1} \Bigr]^a{}_b.
\end{split}
\ee
Since 
$\det( I \pm \sqrt{\beta} H) \, [ ( I \pm \sqrt{\beta} H)^{-1}]^a{}_b$
is a cofactor of the matrix $ I \pm \sqrt{\beta} H$,
\eqref{GFK} is indeed a polynomial of $\beta$
of degree $[(D-1)/2]$.

For simplicity, let us define a matrix $S(\beta)$ by
\be
S(\beta):= ( I + \beta Q)^{-1}.
\ee
Using \eqref{CCKY}, we have
\be
\nabla_a \det\!{}^{1/2} ( I + \beta Q)
= - 2 \beta \xi_d
\Bigl[ H S(\beta) \Bigr]^d{}_a
\, \det\!{}^{1/2} ( I + \beta Q ),
\ee
\be
\begin{split}
\nabla_a S_{bc}(\beta) 
&= \beta S_{ba}(\beta)
\xi^d \Bigl[ H S(\beta) \Bigr]_{dc} 
 - \beta S_{bd}(\beta)  \xi^d
\Bigl[ H S(\beta) \Bigr]_{ac} \cr
& + \beta \Bigl[ H S(\beta) \Bigr]_{ba}
\xi^d S_{dc}(\beta) 
 - \beta \Bigl[ H S(\beta) \Bigr]_{bd} \xi^d
S_{ac}(\beta).
\end{split}
\ee
Combining these relations, we have
\bel{NKB}
\nabla_a K_{bc}( \beta)
= \det\!{}^{1/2}( I + \beta Q) \, \xi^d
X_{abc;d}(\beta),
\ee
where
\be
\begin{split}
X_{abc;d}(\beta)
&= 2 \beta \Bigl[ H S(\beta) \Bigr]_{ad}
S_{bc}(\beta) 
-  \beta \Bigl[ H S(\beta) \Bigr]_{bd}
S_{ca}(\beta) 
-  \beta \Bigl[ H S(\beta) \Bigr]_{cd}
S_{ab}(\beta) \cr
& \ \ + \beta S_{bd}(\beta)
\Bigl[ H S(\beta) \Bigr]_{ca} 
+ \beta S_{cd}(\beta)
\Bigl[ H S(\beta) \Bigr]_{ba}.
\end{split}
\ee
Then with help of \eqref{NKB}, it is easy to
check that the following relations hold:
\be
\nabla_{(a} K_{bc)}(\beta) = 0.
\ee
Therefore we have
\be
\nabla_{(a} K^{(j)}_{bc)} = 0.
\ee

\textit{Proof of Theorem $1$}. In terms of generating function,
Theorem $1$ \eqref{Thm1} can be written as follows
\bel{EQ1}
K_{e(a}(\beta_1) \nabla^e K_{bc)}(\beta_2)
-K_{e(a}(\beta_2) \nabla^e K_{bc)}(\beta_1)
= 0.
\ee
Let
\be
F_{abc}(\beta_1, \beta_2):=
\frac{K_{ea}(\beta_1) \nabla^e K_{bc}(\beta_2)}
{\det\!{}^{1/2} ( I + \beta_1 Q)
\det\!{}^{1/2} ( I + \beta_2 Q)}. 
\ee
\eqref{EQ1} is equivalent to
\be
F_{(abc)}(\beta_1, \beta_2) - F_{(abc)}(\beta_2, \beta_1) = 0.
\ee
Using the explicit form of $\nabla^e K_{bc}(\beta_2)$,
we have
\be
\begin{split}
F_{abc}(\beta_1, \beta_2)
&= \beta_2 \xi^d S_{ea}(\beta_1) \cr
& \times \Bigl(
2 \bigl[ H S(\beta_2) \bigr]_{ed} S_{bc}(\beta_2)
- \bigl[ H S(\beta_2) \bigr]_{bd} S_c{}^e(\beta_2)
- \bigl[ H S(\beta_2) \bigr]_{cd} S^e{}_b(\beta_2) \cr
& \qq
+ S_{bd}(\beta_2) \bigl[ H S(\beta_2) \bigr]_c{}^e
+ S_{cd}(\beta_2) \bigl[ H S(\beta_2) \bigr]_b{}^e \Bigr) \cr
&= \beta_2 \xi^d
\Bigl( 2 \bigl[ H S(\beta_1) S(\beta_2) \bigr]_{ad}
S_{bc}(\beta_2) \cr
& \qq
- \bigl[ H S(\beta_2) \bigr]_{bd} 
\bigl[ S(\beta_1) S(\beta_2) \bigr]_{ca}
- \bigl[ H S(\beta_2) \bigr]_{cd}
\bigl[ S(\beta_1) S(\beta_2) \bigr]_{ab} \cr
& \qq
+ S_{bd}(\beta_2) 
\bigl[ H S(\beta_1) S(\beta_2) \bigr]_{ca}
+ S_{cd}(\beta_2)
\bigl[ H S(\beta_1) S(\beta_2) \bigr]_{ba}
\Bigr). 
\end{split}
\ee
Then
\be
F_{(abc)}(\beta_1, \beta_2)
= 2 \beta_2 \xi^d
\Bigl( 
S_{(bc}(\beta_2) 
\bigl[ H S(\beta_1) S(\beta_2) \bigr]_{a)d} 
-
\bigl[ S(\beta_1) S(\beta_2) \bigr]_{(bc}
\bigl[ H S(\beta_2) \bigr]_{a)d} \Bigr).
\ee
Note that
\bel{BSBS}
\beta_2 S(\beta_2) - \beta_1 S(\beta_1)
= (\beta_2 - \beta_1) S(\beta_1) S(\beta_2).
\ee
Then
\[
\begin{split}
& F_{(abc)}(\beta_1, \beta_2)
- F_{(abc)}(\beta_2, \beta_1) \cr
&= 2 (\beta_2 - \beta_1)
\xi^d
\Bigl( 
\bigl[ S(\beta_1) S(\beta_2) \bigr]_{(bc}
\bigl[ H S(\beta_1) S(\beta_2) ]_{a)d}
- \bigl[ S(\beta_1) S(\beta_2) \bigr]_{(bc}
\bigl[ H S(\beta_1) S(\beta_2) \bigr]_{a)d}
\Bigr) \cr
&= 0.
\end{split}
\]
This completes the proof of Theorem $1$.
\hfill $\square$

Let $\eta_a(\beta)$ be the generating function of $\eta^{(j)}_a$:
\bel{GFE}
\eta_a(\beta) = \sum_{j=0}^{k-1} \eta_a^{(j)} \beta^j
= K_{ab}(\beta) \xi^b.
\ee

\textit{Proof of Theorem $2$}.  In terms of the generating function \eqref{GFE}, 
the theorem $2$ is equivalent to
\be
\mathcal{L}_{\eta(\beta)} h_{ab} = 0.
\ee
The left-handed side is
\bel{Leb}
\mathcal{L}_{\eta(\beta)} h_{ab}
= \eta^c(\beta) \nabla_c h_{ab} + h_{cb} \nabla_a \eta^c(\beta)
+ h_{ac} \nabla_b \eta^c(\beta).
\ee
Using \eqref{CCKY}, the first term in the right-handed side of \eqref{Leb} becomes
\bel{LH1}
\eta^c(\beta) \nabla_c h_{ab}
= \xi_b \eta_a(\beta) - \xi_a \eta_b(\beta).
\ee
Let us examine the second and third terms.  
\be
\begin{split}
U_{ab}(\beta)&:= h_{cb} \nabla_a \eta^c(\beta)
+ h_{ac} \nabla_b \eta^c(\beta) \cr
&= h_{cb} \nabla_a \bigl( K^c{}_d(\beta) \xi^d \bigr)
+ h_{ac} \nabla_b \bigl( K^c{}_d(\beta) \xi^d \bigr) \cr
&= \bigl[ K(\beta) H \bigr]_{db} \nabla_a \xi^d
+ \bigl[ K(\beta) H \bigr]_{ad} \nabla_b \xi^d 
+ \xi^d \bigl( h_{cb} \nabla_a K^c{}_d(\beta)
+ h_{ac} \nabla_b K^c{}_d(\beta) \bigr).
\end{split}
\ee
Note that
\be
\bigl[ K(\beta) H \bigr]_{db} \nabla_a \xi^d
+ \bigl[ K(\beta) H \bigr]_{ad} \nabla_b \xi^d
= \mathcal{L}_{\xi} \bigl[ K(\beta) H \bigr]_{ab} 
- \nabla_{\xi} \bigl[ K(\beta) H \bigr]_{ab} = 0.
\ee
Here we have used \eqref{LieD} and \eqref{CovD}.

Let
\be
V_{ab}(\beta):= \frac{\xi^d h_{ac} \nabla_b K^c{}_d(\beta)}
{\det\!{}^{1/2}( I + \beta Q) }.
\ee
Then
\be
U_{ab}(\beta)= \det\!{}^{1/2}( I + \beta Q)
\Bigl( V_{ab}(\beta) - V_{ba}(\beta)\Bigr)
= 2 \det\!{}^{1/2}( I + \beta Q)\,
V_{[ab]}(\beta). 
\ee
Using \eqref{NKB}, we have
\be
V_{ab}(\beta)
= \beta \xi^d \xi^f
\Bigl\{ [ H S(\beta) ]_{ad} [ HS(\beta) ]_{bf} - S_{df} [ QS(\beta) ]_{ab}
+ [ QS(\beta) ]_{ad} S_{bf}(\beta) \Bigr\},
\ee
\be
2 V_{[ab]}(\beta)
= \beta \xi^d \xi^f
\Bigl\{
[ Q S(\beta) ]_{ad} S_{bf}(\beta) - S_{ad}(\beta) [ Q S(\beta) ]_{bf}
\Bigr\}.
\ee
Note that
\be
\beta Q S(\beta) = I - S(\beta).
\ee
Then
\be
2 V_{[ab]}(\beta)
= \beta \xi^d \xi^f 
\Bigl\{ g_{ad} S_{bf}(\beta) - S_{ad}(\beta) g_{bf} \bigr\}
= \xi_a S_{bf}(\beta) \xi^f - \xi_b S_{ad}(\beta) \xi^d.
\ee
Therefore
\bel{LH2}
U_{ab}(\beta) = \xi_a \eta_b(\beta) - \xi_b \eta_a(\beta).
\ee
Adding \eqref{LH1} and \eqref{LH2}, we have
\be
\mathcal{L}_{\eta(\beta)} h_{ab} = 0.
\ee
This completes the proof of Theorem $2$.
\hfill $\square$

The first relation of Corollary is equivalent to
\bel{T2a}
\mathcal{L}_{\eta^{(i)}} K^{(j)} = 0,
\ee
which immediately follows from Theorem $2$.

The second relation of Corollary is equivalent to
\be
\mathcal{L}_{\eta^{(i)}} \eta^{(j)} = 0.
\ee
Note that
\bel{LDxi}
\mathcal{L}_{\xi} \xi = [ \xi, \xi ] = 0,
\ee
\be
\begin{split}
\mathcal{L}_{\xi} \eta^{(j)a} &= \mathcal{L}_{\xi}
( K^{(j)a}{}_b \xi^b) \cr
& = (\mathcal{L}_{\xi} K^{(j)a}{}_b) \xi^b + K^{(j)a}{}_b
( \mathcal{L}_{\xi} \xi^b ) \cr
&= 0.
\end{split}
\ee
Here we have used \eqref{LieD} and \eqref{LDxi}.
Then
\be
\mathcal{L}_{\eta^{(j)}} \xi = [ \eta^{(j)}, \xi ] = - \mathcal{L}_{\xi} \eta^{(j)} = 0.
\ee
Now, using this relation and \eqref{T2a}, we easily see that
\be
\begin{split}
\mathcal{L}_{\eta^{(i)}} \eta^{(j)a}
&= \mathcal{L}_{\eta^{(i)}} ( K^{(j)a}{}_b \xi^b ) \cr
&= ( \mathcal{L}_{\eta^{(i)}} K^{(j)a}{}_b) \xi^b
+ K^{(j)a}{}_b ( \mathcal{L}_{\eta^{(i)}} \xi^b ) \cr
&= 0.
\end{split}
\ee
This completes the proof of Corollary.


\section{Separation of variables in the Hamilton-Jacobi equation}

A geometric characterisation of the separation of variables 
in the geodesic Hamilton-Jacobi equation was given 
by Benenti-Francaviglia \cite{BF} and 
Kalnins-Miller \cite{KM}. Here, we use
the following result in \cite{KM}.\\
{\bf{Theorem}} Suppose there exists a $N$-dimensional vector space $\mathcal{A}$
of rank-2
Killing tensors on $D$-dimensional space $(M,g)$. 
Then the geodesic Hamilton-Jacobi 
equation has a separable coordinate system if and 
only if the following conditions hold\footnote{We put $n_2=0$ 
for theorem $4$ in \cite{KM}. 
This condition is satisfied in the case of a positive 
definite metric $g$.}:
\begin{enumerate}
\item[(1)] $[A,B]_S=0$ for each $A$,$B$ $\in \mathcal{A}$.
\item[(2)] There exist $(D-n)$-independent simultaneous eigenvectors $X^{(a)}$ 
for every $A \in \mathcal{A}$.
\item[(3)] There exist $n$-independent commuting Killing vectors $Y^{(\alpha)}$.
\item[(4)] $[A,Y^{(\alpha)}]_S=0$ for each $A \in \mathcal{A}$.
\item[(5)] $N=(2 D+n^2-n)/2$.
\item[(6)] $g(X^{(a)}, X^{(b)})=0$ if $1 \le a < b \le D-n$, \\and $g(X^{(a)}, Y^{(\alpha)})=0$
for $1 \le a \le D-n,~D-n+1 \le \alpha \le D$.
\end{enumerate}
We assume that the Killing tensors $K^{(j)}$ and
$K^{(ij)}=\eta^{(i)} \otimes \eta^{(j)}+\eta^{(j)} \otimes \eta^{(i)}$ given in section 2
form a basis for $\mathcal{A}$.
Note that in the odd dimensional case the last Killing Yano tensor
$f^{(k-1)}$ is a Killing vector, and hence the corresponding Killing tensor 
$K^{(k-1)} \propto f^{(k-1)} f^{(k-1)}$ is reducible \cite{KKPF}.
Then, it is easy to see that the conditions $(1) \sim (6)$ hold.
Indeed, the relation $K^{(i)} K^{(j)} = K^{(j)} K^{(i)}$ implies that there exist
simultaneous eigenvectors $X^{(a)}$ for $K^{(i)}$ satisfying conditions (2) and (6). 
Other conditions are direct consequences of Theorem 1 and Corollary.\\


\section{Example}

Finally we describe the Kerr-NUT de Sitter metric as an example, 
which was fully studied in \cite{CLP,HHOY,FK,KF,PKVK,FKK,KKPF,KKPV}.
The $D$-dimensional metric takes the form \cite{CLP}:\\
\noindent
(a)~$D=2n$
\begin{equation}
g=\sum_{\mu=1}^{n}\frac{dx_{\mu}^2}{Q_{\mu}}+\sum_{\mu=1}^{n} Q_{\mu}
\left( \sum_{k=0}^{n-1} A^{(k)}_{\mu} d\psi_k \right)^2
\end{equation}
\noindent
(b)~$D=2n+1$
\begin{equation}
g=\sum_{\mu=1}^{n}\frac{dx_{\mu}^2}{Q_{\mu}}+\sum_{\mu=1}^{n} Q_{\mu}
\left( \sum_{k=0}^{n-1} A^{(k)}_{\mu} d\psi_k \right)^2+ S
\left( \sum_{k=0}^{n} A^{(k)} d\psi_k \right)^2
\end{equation}
The functions $Q_{\mu}$ are given by
\begin{equation}
Q_{\mu}=\frac{X_{\mu}}{U_{\mu}},~~U_{\mu}
=\prod_{\stackrel{\scr \nu=1}{(\nu \ne \mu)}}^n (x_{\mu}^2-x_{\nu}^2),
\end{equation}
where $X_{\mu}$ is a function depending only on $x_{\mu}$  and
\begin{equation}
A^{(k)}_{\mu}
=\sum_{\stackrel{\scr 1 \le \nu_1 < \cdots < \nu_k \le n}{ (\nu_i \ne \mu)}} 
x_{\nu_1}^2x_{\nu_2}^2
\cdots x_{\nu_k}^2,~~
A^{(k)}=\sum_{1 \le \nu_1 < \cdots < \nu_k \le n } x_{\nu_1}^2x_{\nu_2}^2
\cdots x_{\nu_k}^2,~~S=\frac{c}{A^{(n)}}
\end{equation}
with a constant $c$. The CKY tensor is written as \cite{KF}
\begin{equation}
h=\frac{1}{2}\sum_{k=0}^{n-1} dA^{(k+1)} \wedge d\psi_k
\end{equation}
with the associated vector $\xi=\partial/\partial \psi_0$.
The assumptions $(a1), (a2)$ and $(a3)$ are clearly satisfied.
The commuting Killing tensors $K^{(j)}$ and Killing vectors $\eta^{(j)}$
are calculated as \cite{KF,PKVK}
\begin{eqnarray}
K^{(j)}&=&\sum_{\mu=1}^{n} A^{(j)}_{\mu} (e^{\mu}e^{\mu}+e^{\mu+n}e^{\mu+n})
+\epsilon A^{(j)}e^{2n+1} e^{2n+1},\\
\eta^{(j)}&=& \frac{\partial}{\partial \psi_j},
\end{eqnarray}
where $\epsilon=0$ for $D=2n$ and 1 for $D=2n+1$. 
The 1-forms $\{e^{\mu},~e^{\mu+n},~e^{2n+1} \}$ are
orthonormal bases defined by
\begin{equation}
e^{\mu}=\frac{dx_{\mu}}{\sqrt{Q_{\mu}}},
~~e^{\mu+n}=\sqrt{Q_{\mu}}\left( \sum_{k=0}^{n-1} A^{(k)}_{\mu} d\psi_k \right),~~
e^{2n+1}=\sqrt{S}
\left( \sum_{k=0}^{n} A^{(k)} d\psi_k \right).
\end{equation}


{\bf{Acknowledgements}}

This work is supported by the 21 COE program
``Construction of wide-angle mathematical basis focused on knots".
The work of Y.Y is supported by the Grant-in Aid for Scientific
Research (No. 19540304 and No. 19540098)
from Japan Ministry of Education. 
The work of T.O is supported by the Grant-in Aid for Scientific
Research (No. 18540285 and No. 19540304)
from Japan Ministry of Education.


\appendix


\section{Generating function of $K^{(j)}_{ab}$}

In this appendix, we rederive the expression of the generating
function of $K^{(j)}$ directly from the definition \eqref{DKj}.

\subsection{Auxiliary operators}

It is convenient to
introduce auxiliary fermionic creation/annihilation operators:
\be
\bar{\psi}^a, \qq
\psi_a, \qq
a=1,2,\dotsc, D
\ee
such that
\be
\{ \psi_a, \psi_b \} = 0, \qq
\{ \bar{\psi}^a, \bar{\psi}^b \} = 0, \qq
\{ \psi_a, \bar{\psi}^b \} = \delta^b_a.
\ee
Also let
\be
\bar{\psi}_a:= g_{ab} \bar{\psi}^b, \qq
\psi^a:= g^{ab} \psi_b.
\ee
\be
\{ \psi_a, \bar{\psi}_b \} = g_{ab}, \qq
\{ \psi^a, \bar{\psi}^b \} = g^{ab}.
\ee
The Fock vacuum is defined by
\be
\psi_a| 0 \rangle = 0, \qq
\langle 0 | \bar{\psi}^a = 0, \qq
a=1,2,\dotsc, D,
\ee
with a normalization
\be
\langle 0 | 0 \rangle = 1.
\ee
To a $2$-form $h$
\be
h= \frac{1}{2} h_{ab} \de x^a \wedge \de x^b,
\ee
let us associate the following operators:
\be
h_{\bar{\psi}} := \frac{1}{2} h_{ab} \bar{\psi}^a \bar{\psi}^b,
\ee
\be
h_{\psi}:= \frac{1}{2} h^{ab} \psi_a \psi_b.
\ee
Note that
\be
(h_{\bar{\psi}})^j = \frac{1}{(2j)!}
h^{(j)}_{a_1 \dotsm a_{2j}}
\bar{\psi}^{a_1} \dotsm \bar{\psi}^{a_{2j}}.
\ee
\be
\begin{split}
h^{(j)}_{a_1 \dots a_{2j}}
&= \langle 0 | \psi_{a_{2j}} \dotsm \psi_{a_1}
(h_{\bar{\psi}})^j | 0 \rangle \cr
&= (-1)^j \langle 0 | \psi_{a_1} \dotsm \psi_{a_{2j}}
(h_{\bar{\psi}})^j | 0 \rangle.
\end{split}
\ee

\subsection{The generating function of $A^{(j)}$}

Let
\bel{Aj}
\begin{split}
A^{(j)}&:=\frac{1}{(2j)!(j!)^2} 
( h^{(j)}_{c_1 \dotsm c_{2j}} h^{(j)c_1 \dotsm c_{2j}}) \cr
&= \frac{(2j)!}{(2^j j!)^2}
h^{[a_1 b_1} \dotsm h^{a_j b_j]}
h_{[a_1 b_1} \dotsm h_{a_j b_j]}.
\end{split}
\ee
$A^{(j)}$ is nontrivial for $j=0,1,\dotsc, [D/2]$.

Note that
\be
\begin{split}
A^{(j)} &= \frac{1}{(2j)!(j!)^2} h^{(j)}_{c_1 \dotsm c_{2j}}
h^{(j) c_1 \dotsm c_{2j}} \cr
&= \frac{1}{(2j)!(j!)^2 } h^{(j) c_1 \dotsm c_{2j}}
\times (-1)^j \langle 0 | 
\psi_{c_1} \dotsm \psi_{c_{2j}} ( h_{\bar{\psi}})^j | 0 \rangle \cr
&= (-1)^j \langle 0 | \frac{( h_{\psi})^j}{j!} 
\frac{( h_{\bar{\psi}})^j}{j!} | 0 \rangle.
\end{split}
\ee
Then we have
\be
\sum_{j=0}^{[D/2]} A^{(j)} \beta^j
= \langle 0 | \ex^{- \sqrt{\beta} h_{\psi}}
\ex^{\sqrt{\beta} h_{\bar{\psi}}} | 0 \rangle.
\ee
Let us introduce the vielbein
\be
g_{ab} = \delta_{ij} e^i{}_a e^j{}_b.
\ee
(We assume the Euclidean signature.)

Let $E$ be the matrix with elements
\be
E^i{}_a = e^i{}_a.
\ee
Then
\be
H^a{}_b = (E^{-1})^a{}_i \tilde{H}_{ij} E^j{}_b,
\qq
\tilde{H}_{ij} = - \tilde{H}_{ji}.
\ee
Also let
\be
\theta^i = e^i{}_{a} \psi^a, \qq
\bar{\theta}^i = e^i{}_a \bar{\psi}^a, \qq
i=1,2,\dotsc, D.
\ee
Then we have $\theta_i = \theta^i$, $\bar{\theta}_i = \bar{\theta}^i$, and
\be
\{ \theta_i, \theta_j \} = 0, \qq
\{ \bar{\theta}_i, \bar{\theta}_j \} = 0, \qq
\{ \theta_i, \bar{\theta}_j \} = \delta_{ij},
\ee
for $i,j=1,2,\dotsc, D$.
It is well known that any real antisymmetric matrix can be
block diagonalized by some orthogonal matrix.
Therefore, we can choose the vielbein such that $\tilde{H}$
has a block diagonal form and
\be
h_{\psi} = \sum_{\mu=1}^n \lambda_{\mu} \theta_{\mu} \theta_{n+\mu}, \qq
h_{\bar{\psi}} = \sum_{\mu=1}^n \lambda_{\mu} \bar{\theta}_{\mu} 
\bar{\theta}_{n+\mu},
\ee
for $n=[D/2]$. Here we assume that $\lambda_{\mu} \neq 0$.
Note that
\be
E Q E^{-1} = \mathrm{diag}(\lambda_1^2, \lambda_2^2, \dotsc, \lambda_n^2,
\lambda_1^2, \lambda_2^2, \dotsc).
\ee
For odd $D$, the last diagonal entry equals to zero.

Then
\be
\begin{split}
\langle 0| \ex^{-\sqrt{\beta} h_{\psi}} \ex^{\sqrt{\beta} h_{\bar{\psi}} } | 0 \rangle
&= \langle 0 | \prod_{\mu=1}^n ( 1 - \sqrt{\beta} \lambda_{\mu}
\theta_{\mu} \theta_{n+\mu} )
( 1 + \sqrt{\beta} \lambda_{\mu} \bar{\theta}_{\mu}
\bar{\theta}_{n+\mu} ) | 0 \rangle \cr
&= \prod_{\mu=1}^n ( 1 + \beta \lambda_{\mu}^2 ) \cr
&= \det\!{}^{1/2}( I + \beta Q ).
\end{split}
\ee
Here $I$ is the $D\times D$ identity matrix.

We have the generating function of $A^{(j)}$:
\bel{GFA}
\sum_{j=0}^{[D/2]} A^{(j)} \beta^j
 = \det\!{}^{1/2}( I + \beta Q)
= \det( I + \sqrt{\beta} H) = \det( I - \sqrt{\beta} H).
\ee

\subsection{Recursion relations for $K^{(j)}$}

The Levi-Civita tensor satisfies
\bel{LCid}
\varepsilon^{a_1 \dotsm a_r c_1 \dotsm c_{D-r}}
\varepsilon_{b_1 \dotsm b_r c_1 \dotsm c_{D-r}}
= r! (D-r)! \delta^{[a_1}_{b_1} \dotsm \delta^{a_r]}_{b_r}.
\ee
Using \eqref{LCid}, we can check that $K^{(j)}_{ab}$
has the following form:
\bel{Kj}
K^{(j)}_{ab}= A^{(j)} g_{ab}
+ \frac{1}{(2j-1)!(j!)^2}
h^{(j)}_{a c_1 \dotsm c_{2j-1}}
h^{(j) c_1 \dotsm c_{2j-1}}{}_b.
\ee
Here $A^{(j)}$ is defined by \eqref{Aj}.

It is possible to show that
\bel{Th1}
\frac{1}{(2j-1)!(j!)^2}
h^{(j)}_{a c_1 \dotsm c_{2j-1}}
h^{(j) c_1 \dotsm c_{2j-1}}{}_b
= h_{a}{}^{c} K^{(j-1)}_{cd} h^{d}{}_{b}.
\ee
In the matrix notation,
$K^{(j)}$ satisfies the following recursion relation:
\be
K^{(j)} = A^{(j)} I + H K^{(j-1)} H.
\ee
Therefore, we can see that $K^{(j)}$ commutes with $H$.
Thus
\bel{RRK}
K^{(j)} = A^{(j)} I - Q K^{(j-1)}.
\ee
With the initial condition
\be
K^{(0)} = I, \qq
K^{(0)}_{ab} = g_{ab},
\ee
we easily find that
\bel{KAQ}
K^{(j)} = \sum_{l=0}^j (-1)^l A^{(j-l)} Q^l,
\ee
or
\be
K^{(j)a}{}_b = \sum_{l=0}^j (-1)^l A^{(j-l)} ( Q^l)^a{}_b.
\ee
We immediately see that 
\be
K^{(i)} K^{(j)} = K^{(j)} K^{(i)}.
\ee

Using \eqref{GFA}, we can see that $K^{(k)} = 0$
for $k=[(D+1)/2]$. Indeed, by setting $\beta= - x^{-1}$,
\be
\sum_{j=0}^{[D/2]} (-1)^j A^{(j)} x^{-j} = \det\!{}^{1/2}( I - x^{-1} Q )
= x^{-D/2} \det\!{}^{1/2} ( x I - Q ).
\ee
For $D=2k$,
\be
\sum_{j=0}^k (-1)^{k-j} A^{(j)} x^{k-j} = (-1)^k \det\!{}^{1/2} ( x I - Q ).
\ee
If we set $x$ to be an eigenvalue of $Q$, the R.H.S. becomes zero.
Therefore, we can see that
\be
K^{(k)}= \sum_{l=0}^k (-1)^l A^{(k-l)} Q^l = 0, \qq \mbox{ for } \, D=2k.
\ee

Similarly, for $D=2k-1$,
\be
\sum_{j=0}^{k-1} (-1)^{k-j} A^{(j)} x^{k-j} = (-1)^k x^{1/2}
\det\!{}^{1/2}( x I - Q).
\ee
Thus
\be
K^{(k)} = \sum_{l=1}^{k} (-1)^l A^{(k-l)} Q^l = 0, \qq
\mbox{ for }\, D=2k-1.
\ee
Also note that $A^{(j)} = 0$ for $ j \geq [D/2] +1$.
Therefore the recursion relations \eqref{RRK} becomes trivial
for $j \geq k+1$ and $K_{ab}^{(j)}=0$ for $j \geq k$.
$K^{(j)}$ can be written as \eqref{KAQ} for all $j \geq 0$
but 
are nontrivial only for $j=0,1,\dotsc, k-1$.
 
Using \eqref{KAQ} and \eqref{GFA},
we can see that the generating function of $K^{(j)}$
is
\be
K(\beta):= \sum_{j=0}^{k-1} K^{(j)} \beta^j
= \det\!{}^{1/2}( I + \beta Q)\,  ( I + \beta Q)^{-1}.
\ee

\subsection{Proof of \eqref{Th1} }

The L.H.S. of \eqref{Th1} is
\be
\begin{split}
& \frac{1}{(2j-1)!(j!)^2} h^{(j)}_{ac_1 \dotsm c_{2j-1}}
h^{(j) c_1 \dotsm c_{2j-1}}{}_b  \cr
&= \frac{1}{(2j-1)!(j!)^2} h^{(j)c_1 \dotsm c_{2j-1}}{}_b
\times (-1)^j \langle 0 | \psi_a \psi_{c_1} \dotsm \psi_{c_{2j-1}}
( h_{\bar{\psi}})^j | 0 \rangle \cr
&= \frac{(-1)^{j-1}}{(2j-1)!(j!)^2}
h^{(j) c_1 \dotsm c_{2j-1}}{}_b
\langle 0 | \psi_{c_1} \dotsm \psi_{c_{2j-1}} \psi_a
( h_{\bar{\psi}})^j | 0 \rangle \cr
&= \frac{(-1)^{j-1}}{(2j)!(j!)^2}
h^{(j)c_1 \dotsm c_{2j}}
\langle 0 | \psi_{c_1} \dotsm \psi_{c_{2j}} \bar{\psi}_b
\psi_a ( h_{\bar{\psi}} )^j | 0 \rangle \cr
&=(-1)^{j-1}
\langle 0 | 
\frac{( h_{\psi})^j}{j!} \bar{\psi}_b \psi_a 
\frac{( h_{\bar{\psi}})^j}{j!}
| 0 \rangle.
\end{split}
\ee
Then
\be
\begin{split}
K^{(j)}_{ab} &= (-1)^j g_{ab} \langle 0 | 
\frac{( h_{\psi})^j}{j!}
\frac{( h_{\bar{\psi}})^j}{j!} | 0 \rangle 
 - (-1)^j \langle 0 | 
\frac{( h_{\psi})^j}{j!} \bar{\psi}_b
\psi_a \frac{( h_{\bar{\psi}})^j}{j!} | 0 \rangle \cr
&= (-1)^j
\langle 0 | \frac{( h_{\psi})^j}{j!}
\left[ \{ \psi_a, \bar{\psi}_b\} - \bar{\psi}_b \psi_a \right]
\frac{( h_{\bar{\psi}})^j}{j!} | 0 \rangle \cr
&= (-1)^j \langle 0 | 
\frac{( h_{\psi})^j}{j!} \psi_a \bar{\psi}_b
\frac{( h_{\bar{\psi}})^j}{j!} | 0 \rangle.
\end{split}
\ee
Thus
\bel{Kab1}
K^{(j)}_{ab}
= (-1)^j 
 \langle 0 | \frac{( h_{\psi})^j}{j!} \psi_a \bar{\psi}_b
\frac{( h_{\bar{\psi}})^j}{j!} | 0 \rangle.
\ee
Note that
\be
[ \psi_a, h_{\bar{\psi}}] = h_{aa'} \bar{\psi}^{a'},
\ee
\be
\psi_a ( h_{\bar{\psi}})^j | 0 \rangle
= j h_a{}^{a'} \bar{\psi}_{a'} ( h_{\bar{\psi}})^{j-1} | 0 \rangle,
\ee
\be
[ h_{\psi}, \bar{\psi}_b ]
= \psi_{b'} h^{b'}{}_b,
\ee
\be
\langle 0 | ( h_{\psi})^j \bar{\psi}_b
= j \langle 0 | ( h_{\psi})^{j-1} \psi_{b'} h^{b'}{}_b.
\ee
Then 
\be
\begin{split}
(\mbox{L.H.S. of \eqref{Th1}})
&= \frac{1}{(2j-1)!(j!)^2}
h^{(j)}_{a c_1 \dotsm c_{2j-1}} h^{(j) c_1 \dotsm c_{2j-1}}{}_b \cr
&= (-1)^{j-1} \langle 0 | 
\frac{( h_{\psi})^j}{j!} \bar{\psi}_b
\psi_a \frac{( h_{\bar{\psi}})^j}{j!} | 0 \rangle \cr
&= h_a{}^{a'}
(-1)^{j-1}
\langle 0 | 
\frac{( h_{\psi})^{j-1}}{(j-1)!} \psi_{b'} \bar{\psi}_{a'}
\frac{( h_{\bar{\psi}})^{j-1}}{(j-1)!} |0 \rangle h^{b'}{}_{b} \cr
&= h_{a}{}^{a'} K^{(j-1)}_{a' b'} h^{b'}{}_b \cr
&= (\mbox{R.H.S. of \eqref{Th1}}).
\end{split}
\ee
This completes the proof of \eqref{Th1}.


\end{document}